\documentclass[11pt,a4paper]{article}

\usepackage[utf8]{inputenc}
\usepackage[T1]{fontenc}
\usepackage[expansion=false]{microtype}
\usepackage{amsmath,amssymb,amsthm}
\usepackage{booktabs}
\usepackage{array}
\usepackage{tabularx}
\usepackage{enumitem}
\usepackage{xcolor}
\usepackage{authblk}
\usepackage[margin=2.6cm]{geometry}
\usepackage[colorlinks=true,linkcolor=blue!50!black,citecolor=blue!50!black,urlcolor=blue!50!black]{hyperref}

\theoremstyle{plain}
\newtheorem{theorem}{Theorem}[section]
\newtheorem{lemma}[theorem]{Lemma}

\newtheorem{proposition}[theorem]{Proposition}
\theoremstyle{definition}
\newtheorem{definition}[theorem]{Definition}

\theoremstyle{remark}

\newcommand{\Adv}{\mathrm{Adv}}

\newcommand{\DLPS}{\mathrm{DLP\text{-}Sign}}

\newcommand{\EUF}{\mathrm{EUF\text{-}CMA}}
\newcommand{\AIPRF}{\mathrm{AI\text{-}PRF}}
\newcommand{\Astar}{A^{*}}
\newcommand{\LRPRFQ}{\mathrm{LR\text{-}PRF}_{Q}}
\newcommand{\IC}{\mathrm{IC}}

\newcommand{\Z}{\mathbb{Z}}
\newcommand{\F}{\mathbb{F}}
\newcommand{\bits}{\{0,1\}}
\DeclareRobustCommand{\code}[1]{\ifmmode\text{\ttfamily\def\_{\textunderscore\allowbreak}#1}\else{\ttfamily\def\_{\textunderscore\allowbreak}#1}\fi}
\newcommand{\BE}{\mathrm{BE}_{32}}
\newcommand{\Obs}{\mathrm{Obs}}
\newcommand{\Gen}{\mathrm{Gen}}
\newcommand{\Init}{\mathrm{Init}}
\newcommand{\QRat}{\mathsf{QRat}}

\title{From Architectural Traces to Conditional Key-Recovery Bounds for Deterministic secp256k1 ECDSA}

\author{Ilia Prosianoi}
\affil{Department of Computer Science, Higher School of Economics, Moscow, Russia\\}
\date{November 2026}

\begin{document}
\emergencystretch=1.5em
\maketitle

\begin{abstract}
RFC~6979 derives the ECDSA nonce from the signing key $d$ and the 32-byte digest $h$ via HMAC\_DRBG. Earlier work (arXiv:2611.18449) isolated the assumption $A^*$ for multi-query unforgeability on $\mathbb{Z}_n^*$ and reduced $A^*$ in the ideal-compression model to discrete logarithm with a signing oracle. Here we ask when the instruction and memory-address trace $T$ of an AArch64 secp256k1 signer, observed with one low-$s$ signature $(r,s)$ on a fixed digest $h$ under a prior $J$ on $(d,h,z)$---such as $D_j = [2^{j-1},2^j)$, where $H(d\mid Q)=0$ and restriction from $\mathbb{Z}_n^*$ loses $(n-1)/|D_j| \approx 2^{257-j}$---helps recover $d$. For a pinned AArch64 build of Bitcoin Core~v0.14.0, SLEIGH p-code and Z3 proofs partition runs of at most $H$ attempts into $M_H = 2(4^H-1)/3$ trace templates ($3^H-1$ when $h \equiv 0 \pmod n$). Under uniform nonces on secp256k1, $T$ is publicly simulable from $(r,s)$ up to $\mathbb{E}_J[q_X^H] < 1.395\cdot 10^{-77}$ at $H=2$. Relative to signature-free recovery $S_Q^{\mathrm{native}}$, we prove Lean~4 reductions for $b$-bit views of $s$, verification-point views $(Q,r,T_{\mathrm{ver}})$, and two-stage bounded-degree polynomial solvers; meta-reductions and a $(n-1)/(2n)$ liftability test rule out random symmetric coordinate tables. Passing to HMAC\_DRBG reduces to nonce-prefix indistinguishability ($\mathrm{LR\text{-}PRF}_Q$), which under ideal compression reduces to $S_Q^{\mathrm{native,IC}}$ via $256 q_f$ key checks with no signing oracle and no $(n-1)/|D_j|$ loss, and after initialization splits into $2$- and $3$-query HMAC-PRF hybrids. Physical side channels, historical attribution, general full-$s$ native reductions, and fixed-SHA-256 initialization bounds remain open.
\end{abstract}

\section{Introduction}\label{sec:intro}

Deterministic ECDSA~\cite{rfc6979} computes the per-signature nonce from the signing key $d \in [1,n-1]$ and the 32-byte digest $h$ with HMAC\_DRBG. In~\cite{prosianoi2026}, we showed that existential unforgeability of deterministic ECDSA in the standard model rests on signing-transcript indistinguishability $\Astar$ at the public key $Q = dG$: $\Astar$ follows neither from the PRF security of HMAC nor from auxiliary-input PRF security ($\AIPRF$), and with an ideal compression function $\Astar$ reduces to discrete logarithm given a randomized-ECDSA signing oracle ($\DLPS$), dropping that oracle only when the outer message hash is programmed as a random oracle.

Different obstacles arise when a compiled signer runs once on a fixed digest $h$ under a prior $J$ on $(d,h,z)$:
\begin{enumerate}[leftmargin=*,itemsep=2pt]
\item \emph{Control-flow and address traces.} Machine code branches on candidate rejections ($k = 0$, $k \ge n$, $r = 0$, $s_{\mathrm{raw}} = 0$) and low-$s$ normalization ($s_{\mathrm{raw}} > \lfloor n/2\rfloor$). Even when register and memory values remain hidden, does the full program-counter and memory-address trace $T$ reveal more about $d$ than the published signature $(r,s_{\mathrm{low}})$?
\item \emph{Fixed digests and short-key priors.} When $h$ is fixed before signing, no message hash can be programmed. When $d$ is drawn from an interval $D_j = [2^{j-1},2^j)$ of size $2^{j-1} \ll n$, $Q = dG$ determines $d$ uniquely, so $H(d\mid Q) = H_\infty(d\mid Q) = 0$ and information-theoretic extraction fails; at the same time, restricting a bound from uniform $\Z_n^*$ to $D_j$ multiplies the error by $(n-1)/|D_j| \approx 2^{257-j}$ ($2^{117}$ at $j = 140$).
\item \emph{Actual curve coordinates.} Can one-signature key-recovery bounds be proved on secp256k1 itself, with $C(P) = x(P)\bmod n$, relative to signature-free recovery $S_Q^{\mathrm{native}}(J,t')$ under the same prior $J$, rather than under an idealized coordinate table?
\end{enumerate}

\paragraph{Short-key priors and historical signatures.}
Intervals $D_j = [2^{j-1},2^j)$ appear in public discrete-logarithm challenges on secp256k1 such as the Bitcoin Puzzle transactions, where a public key and one spending signature are on chain. Replaying those 14 available signatures shows byte-for-byte agreement across Bitcoin Core~0.14.0--0.16.3 and several independent libraries under matching transaction versions, while the 2015 funding transaction predates Core~0.14.0. Throughout, $D_j$ is treated solely as a prior in a defined recovery game; nothing here attributes an on-chain signature to a specific binary, asserts that a trace was ever recorded, or recovers any live key.

\paragraph{Contributions.}
\begin{enumerate}[leftmargin=*,itemsep=2pt]
\item We define the one-signature key-recovery game under an arbitrary prior $J$ on $(d,h,z)$ with public-key context generation (\code{TimedPubKey\allowbreak ContextGen}), and verify instruction- and p-code-level contracts for a pinned AArch64 build of the Bitcoin Core~v0.14.0 signer (raw seed $\BE(d)\,\|\,h$, retry limit $L = 2^{32}-1$), partitioning all runs of at most $H$ attempts into $M_H = 2(4^H-1)/3$ public trace templates ($3^H-1$ when $e = 0$; Sections~\ref{sec:model}--\ref{sec:machine}).
\item Under uniform nonces on secp256k1, we prove that the architectural trace $T$ is publicly simulable from $(r,s_{\mathrm{low}})$ up to tail $\mathbb{E}_J[q_X^H] < 1.395\cdot 10^{-77}$ at $H=2$ (Theorem~\ref{thm:trace-sim}). Relative to $S_Q^{\mathrm{native}}(J,t')$, we prove Lean~4 reductions for $b$-bit views of $s_{\mathrm{low}}$, verification-point views $(Q,r,T_{\mathrm{ver}})$ ($2$ checks at $b=0$), and two-stage bounded-degree polynomial solvers ($2\,S_Q^{\mathrm{native}}$, with zero root-finding error $\varepsilon_{\mathrm{rf}}=0$ in the affine case $\Delta_S+\Delta_D\le 1$). Against validity-gated black-box solvers, we prove key-oblivious and target-gated meta-reduction bounds and show that random symmetric coordinate tables are distinguishable from secp256k1 with advantage $\ge (n-1)/(2n)$ (Sections~\ref{sec:tracesim}--\ref{sec:native}).
\item We separate four regimes---native secp256k1 with fixed SHA-256, native secp256k1 with an ideal compression function ($\IC$), an opaque generic group with a symmetric coordinate table, and conditional numerical examples---and reduce deterministic signing transfer to nonce-prefix indistinguishability at the public key ($\LRPRFQ$). Under public $r$-filtering (such as low-$R$ grinding), we prove $\eta_N(\mathcal{A}_t) \ll \delta_{\mathrm{sig}} \approx 1/2$ across any solver family with small ideal recovery $\varepsilon_I(\mathcal{A}_t)$. In the $\IC$ model ($H=1,2$), $\LRPRFQ$ reduces to $S_Q^{\mathrm{native,\IC}}$ via $256(q_{f,\mathrm{ctx}}+q_{f,\mathrm{dist}})$ key checks with no signing oracle and no $(n-1)/|D_j|$ loss; for fixed SHA-256, post-initialization rounds reduce to $2$- and $3$-query HMAC-PRF hybrids while the initialization bound $\varepsilon_{\mathrm{init},Q}$ remains open (Sections~\ref{sec:transfer}--\ref{sec:eval}).
\end{enumerate}

All claims are tagged as kernel-checked in Lean~4.8.0 (\code{import Lean} only, zero Mathlib, \code{trustLevel = 0}; declaration map in Appendix~\ref{app:map})~[M-Lean], checked via pinned SLEIGH p-code and Z3~4.15.4~[M-SMT], proved here~[T], established by finite, emulator, or OS-CSPRNG tests~[E], taken from the literature~[L], or conditional/open~[O]. In Lean, \code{\#print axioms} checks logical derivation from a theorem's stated premises, not the external premises themselves.

\section{Related work}\label{sec:related}

\subsection{Relation to \texorpdfstring{\cite{prosianoi2026}}{[1]}}\label{sec:delta-overview}

Sections~3--6 and~8 of~\cite{prosianoi2026} formulated $\Astar$, separated $\Astar$ from $\AIPRF$ in both directions, proved an ideal-compression bound for $\Astar$ relative to $\DLPS$, surveyed 31 secp256k1 libraries (including raw-digest seeding before libsecp256k1~0.2.0 and low-$R$ grinding), and bounded the Fourier coefficients of the low-$R$ nonce set. Those results serve as starting points here. Table~\ref{tab:delta-summary} summarizes how the two settings differ, and Appendix~\ref{app:delta} gives a theorem-by-theorem comparison.

\begin{table}[t]
\centering
\caption{Comparison between~\cite{prosianoi2026} and this work.}\label{tab:delta-summary}
\small
\begin{tabularx}{\textwidth}{@{}>{\raggedright\arraybackslash}p{2.9cm}XX@{}}
\toprule
Aspect & Prosianoi~\cite{prosianoi2026} & This work \\
\midrule
Security goal & Multi-query $\EUF$ unforgeability and transcript indistinguishability $\Astar$ for uniform $d \leftarrow \Z_n^*$ & Verifiable key recovery $p_b S^P_{\mathcal{O},b}(J,t)$ from one signature and architectural trace $T$ under arbitrary priors $J$ (e.g.\ $D_j$) \\
Ideal-compression bound & Thm.~6.1: $\Adv_{\Astar} \le \Adv^{\DLPS} + O(q_s(q_f+q_s)/2^{256})$, using a Rand-ECDSA signing oracle unless outer $H$ is programmed & $\LRPRFQ^{\IC}$ ($H=1,2$): reduces to $S_Q^{\mathrm{native,\IC}}$ on a fixed digest via $256 q_{f,\mathrm{tot}}$ key checks, with no signing oracle and no $n/|D_j|$ factor \\
Auxiliary-input notion & Multi-query chosen-digest $\AIPRF$ vs.\ $\Astar$; low-$R$ grinding preserves $\EUF$ under $\Astar_c$ (Thm.~8.1) & Raw-prefix $\LRPRFQ$ under \code{TimedPubKey\allowbreak ContextGen}; $\eta_N(\mathcal{A}_t) \ll \delta_{\mathrm{sig}}$ under $r$-filtering when $\varepsilon_I(\mathcal{A}_t)\ll 1$ \\
Implementation scope & Source audit of 31 libraries; $h \ge n$ test vectors & Pinned AArch64 Mach-O binaries, SLEIGH p-code + Z3 proofs, Unicorn replay, 13-module Lean~4.8.0 formalization \\
\bottomrule
\end{tabularx}
\end{table}

\subsection{ECDSA reductions, conversion maps, and meta-reductions}

Brown~\cite{brown2005} analyzed ECDSA in generic groups, and Fersch, Kiltz, and Poettering~\cite{fkp2016} decomposed the conversion map into algebraic shells around a bijective random oracle to prove existential unforgeability via forking; their TCC~2017 paper~\cite{fkp2017} studied one-per-message unforgeability of (EC)DSA variants. Proofs that reprogram a message hash or idealize the conversion map as a bijective oracle do not give polynomial-loss reductions for key recovery from one signature on a fixed digest under $C(P)=x(P)\bmod n$.

Hartmann and Kiltz~\cite{hk2023} proved algebraic-group meta-reduction limits on ECDSA unforgeability when the conversion map is non-programmable. Section~\ref{sec:barriers} studies fixed-digest restricted-domain key recovery against representation-invariant validity-gated solvers; Table~\ref{tab:hk-compare} compares the tasks, oracle rules, $Q$-access models, and quantifier orders. Partial-nonce lattice attacks (Boneh--Venkatesan~\cite{bv1996}, Nguyen--Shparlinski~\cite{ns2002,ns2003}, Minerva~\cite{minerva2020}, LadderLeak~\cite{ladderleak2020}) exploit leaked bits of the ephemeral nonce $k$; a short key $d \in D_j$ paired with a 256-bit nonce is a different problem.

\subsection{HMAC-DRBG, extraction, and auxiliary-input security}

Ye et al.~\cite{ye2017} verified in Coq the functional correctness and hybrid pseudorandomness of mbedTLS HMAC-DRBG compiled with CompCert, assuming high-entropy secret seed material. In RFC~6979 and Bitcoin Core~v0.14.0, `Initialize` instead keys HMAC with $K_0 = \mathtt{0x00}^{32}$ and feeds $(d,h)$ as message blocks while $Q = dG$ is public, so $H_\infty(d\mid Q)=0$. Computational unpredictability when $H_\infty(d\mid Q)=0$ belongs to auxiliary-input cryptography (Dodis, Kalai, and Lovett~\cite{dkl2009}; Dodis et al.~\cite{dgk2009}). Bellare's HMAC-PRF bound~\cite{bellare2006} (and dual-PRF bounds~\cite{bbgs2023}) applies to `Generate` once the HMAC key is uniform, whereas Krawczyk's extract-then-expand analysis~\cite{krawczyk2010,rfc5869} frames `Initialize` as a computational-extractor hypothesis that remains open for fixed SHA-256 given $Q = dG$.

\subsection{Binary noninterference, verified cryptography, and Lean formalizations}

Constant-time and relational verifiers such as \code{ct-verif}~\cite{almeida2016} (which allows public-output declassification) and \code{Binsec/Rel}~\cite{daniel2023} check secret-independence of control flow and memory addresses in LLVM IR or machine code. Verified implementations include HACL*~\cite{hacl2017}, Fiat Cryptography~\cite{fiat2019}, verified \code{safegcd} in libsecp256k1~\cite{safegcd2021} (technical post), system-level cryptographic server verification~\cite{server2024}, verified XMSS/XMSSMT implementations linked to security proofs~\cite{barbosa2026}, and machine-code/Lean verification of Signal~\cite{signal2026} (technical report). Here the binary trace is not singleton: the ECDSA retry loop and low-$s$ normalization branch on candidate outcomes, and our goal is to bound the resulting trace distribution in a restricted-domain key-recovery game.

In generic groups, Shoup~\cite{shoup1997} and Mironov, Mityagin, and Nissim~\cite{mmn2006} give the $\Omega(\sqrt{|D_j|})$ lower bound on intervals, confined to opaque group encodings. In Lean~4, VCVio~\cite{vcvio2026} and HOPSCOTCH~\cite{hopscotch2026} formalize monadic oracle games; our Lean~4.8.0 development instead works directly over a finite-support rational type ($\QRat$) using core \code{import Lean} (\code{trustLevel = 0}) for exact finite-horizon probability accounting.

\section{Model and definitions}\label{sec:model}

\subsection{Curve, digests, and retry horizon}

Let $E(\F_p): y^2 = x^3 + 7$ be secp256k1 over $p = 2^{256}-2^{32}-977$, with prime-order subgroup $\langle G\rangle$ of order $n < 2^{256}$ and word space $[0,U)$, $U = 2^{256}$. Write $C : \langle G\rangle \setminus \{\mathcal{O}\} \to \Z_n$ for $C(P) = x(P) \bmod n$. The signing key is $d \in [1,n-1]$ with public key $Q = dG$.

We distinguish the 32-byte digest $h \in \bits^{256}$, its big-endian integer $m = \mathrm{int}(h) \in [0,U)$, and the scalar $e = m \bmod n \in \Z_n$. Given candidates $\mathbf{w} = (w_0, \dots, w_{L-1}) \in [0,U)^L$ up to the 32-bit counter limit $L = 2^{32}-1$, attempt $i$ is accepted iff
\begin{equation}\label{eq:accept}
1 \le w_i < n, \qquad r_i := C(w_i G) \ne 0, \qquad s_{\mathrm{raw},i} := w_i^{-1}(e + r_i d) \bmod n \ne 0 .
\end{equation}
The condition $s_{\mathrm{raw},i} \ne 0$ is equivalent to $eG + r_i Q \ne \mathcal{O}$ and is publicly testable from $(Q,e,w_i)$ without $d$~[M-Lean]. At the first accepted index $i^* < L$, the signer sets $k = w_{i^*}$ and outputs $(r, s_{\mathrm{low}})$ with $s_{\mathrm{low}} = \min(s_{\mathrm{raw}}, n - s_{\mathrm{raw}})$ (event $P$); if all $L$ candidates fail, nothing is output ($\neg P$).

We compare two signer worlds on the same curve, limit $L$, and prior $J$:
\begin{itemize}[leftmargin=*,itemsep=1pt]
\item \emph{Ideal-nonce world ($I$):} $w_0,\dots,w_{L-1} \leftarrow [0,U)^L$ are independent and uniform.
\item \emph{Fixed-SHA world ($F$):} $w_i = \Gen_F(d,h)_i$ is the $i$-th 32-byte output of the Bitcoin Core~v0.14.0 HMAC-SHA-256 DRBG seeded with $S = \BE(d)\,\|\,h$.
\end{itemize}

\subsection{Priors, context generation, and joint vs.\ conditional recovery}

Let $W = (h,z)$ be the public context (digest $h$ and advice $z$), and $J(d,W)$ a finite-support probability distribution on $[1,n-1] \times \mathcal{W}$. For short-key intervals, $d$ is supported on $D_j = [2^{j-1},2^j)$ of size $2^{j-1}$.

\paragraph{Public-key context generation.}
Because $\Gen_F(d,h)$ is deterministic, conditional statistical independence $W \perp \Gen_F(d,h) \mid (d,h)$ holds trivially even when $z = \Gen_F(d,h)_0$, which reveals $d$ from $(r,s_{\mathrm{low}})$ in two curve checks (\code{FixedSHABridge.\allowbreak advice\_leakage\_breaks\_etaN})~[M-Lean]. When $W$ is generated dynamically rather than fixed beforehand, we require \code{TimedPubKey\allowbreak ContextGen}: from $(\iota,d) \sim \mu_0$, a randomized algorithm $\mathcal{G}_{\mathrm{ctx}}(\iota, Q=dG, \omega_W)$ of cost $t_{\mathrm{ctx}}$ outputs $W$ from $(\iota,Q)$ without receiving $d$~[M-Lean].

\paragraph{Joint mass and conditional probability.}
Write $\mathcal{O}_b(d,W,\omega)$ for the observation under world $b \in \{I,F\}$ when $P_b$ occurs---for example $Q$ alone, $r$ with $b$ bits of $s_{\mathrm{low}}$, the full signature $\sigma = (r,s_{\mathrm{low}})$, or $(\sigma,T)$. Let $p_b = \Pr_{J,b}[P_b] = 1 - u_L^b$, where $u_H^b$ is the probability that the first $H$ candidates are rejected. For resource budget $t$ we track both
\begin{align}
p_b S_{\mathcal{O},b}^{P}(J,t) &:= \sup_{A \in \mathcal{B}_t} \Pr_{(d,W)\sim J,\, b}\big[P_b \land A(W,Q,\mathcal{O}_b) = d\big], \label{eq:joint-mass}\\
S_{\mathcal{O},b}^{P}(J,t) &:= \frac{p_b S_{\mathcal{O},b}^{P}(J,t)}{p_b} \qquad (p_b > 0), \label{eq:cond-prob}
\end{align}
against the signature-free baseline $S_Q^{\mathrm{native}}(J,t) := \sup_{M \in \mathcal{B}_t} \Pr_{(d,W)\sim J}[M(W,Q) = d]$. Proving bounds first on the joint mass~\eqref{eq:joint-mass} avoids hidden changes of measure when $p_I \ne p_F$; conditional bounds follow on dividing by a lower bound on $p_b$. Every candidate key $\hat d$ in a reduction is checked via $\hat d G = Q$ (\code{ops.isPubKey}).

\subsection{Architectural trace model \texorpdfstring{$\mathcal{T}_{\mathrm{arch}}$}{T\_arch}}\label{sec:obs-model}

In a fixed public memory layout $\pi$ (stack, context table, output, key, and digest addresses), the trace $T = \Obs(B,d,\pi)$ of binary $B$ is the sequence of events
\[
e_\tau \in \{\,\mathsf{pc}(\mathrm{addr},\mathrm{next}),\ \mathsf{load}(\mathrm{addr},\mathrm{width}),\ \mathsf{store}(\mathrm{addr},\mathrm{width}),\ \mathsf{exit}(\mathrm{addr})\,\},
\]
recording every instruction address, next PC, memory read/write byte address and width, and final return PC. Register and memory values, cycle counts, wall-clock time, cache-line contention beyond byte addresses, branch-predictor or speculative state, power/EM leakage, and multi-execution state are outside $T$. The signature $(r,s_{\mathrm{low}})$ is given publicly alongside $T$.

\section{The compiled AArch64 signer}\label{sec:machine}

\subsection{Pinned binaries and trusted base}

We analyze AArch64 Mach-O binaries compiled from Bitcoin Core~v0.14.0 commit \code{43a987c1\allowbreak cbcd04b9\allowbreak c5905260\allowbreak 6c3c31c5\allowbreak 6abc826e} (32 C/header files pinned in \code{upstream-sha256.json}) with \code{clang -target arm64-apple-macos14.0 -O2 -fno-inline -fno-stack-protector -mgeneral-regs-only -std=c99} (Apple SDK~26.5, build \code{25F70}):
\begin{enumerate}[leftmargin=*,itemsep=1pt]
\item \code{core-v0140.macho} ($66{,}696$ bytes, SHA-256 \code{7f5bb085\ldots}), containing the $200$-byte function \code{\_nonce\_function\_rfc6979} (\code{e75666b5\ldots});
\item \code{drbg.macho} ($33{,}760$ bytes, \code{c7b7e292\ldots}), linking the callback with HMAC-SHA-256 DRBG;
\item \code{retry/signer.macho} ($35{,}440$ bytes, \code{6f1e17fc\ldots}), linking \code{\_secp256k1\_ecdsa\_sign} with field, group, and scalar arithmetic;
\item \code{retry/context.bin} ($65{,}744$ bytes, \code{326df2a9\ldots}), the precomputed signing context table (paired in the 6-file binary manifest with two archived OS-CSPRNG sample files; Section~\ref{sec:eval}).
\end{enumerate}
Instruction lifting uses the Apache-2.0 Ghidra SLEIGH grammar \code{AARCH64:LE:64:v8A} (\code{sleigh\_ref/\allowbreak AARCH64.sla}, SHA-256 \code{ab7a9195\ldots}, checked for p-code agreement against the host \code{pypcode==3.3.3} wheel on all $19{,}462$ 32-bit words of the three Mach-O \code{\_\_text} sections), a symbolic p-code interpreter, Z3~4.15.4, and the Lean~4.8.0 kernel. Python, the SLEIGH specification, and the hardware are part of the trusted base.

\subsection{Binary verification stages}

Five sets of Z3 bit-vector obligations (\code{QF\_ABV}/\code{QF\_AUFBV}) establish:
\begin{enumerate}[leftmargin=*,itemsep=2pt]
\item \emph{Seed layout (\code{arm64/\allowbreak obligations.smt2}):} Under the standard call profile (\code{algo16 = NULL}, \code{ndata = NULL}, \code{counter} $\in [0,L-1]$), a 15-instruction path writes $S = \BE(d)\,\|\,h$ to stack memory with one fixed trace ($M=1$, secret branches $\beta=0$)~[M-SMT][E].
\item \emph{HMAC-SHA-256 DRBG (\code{drbg/\allowbreak check\_drbg.py}, \code{functional/\allowbreak function.smt2}):} Each fixed-counter call to \code{initialize} and \code{generate} has a singleton trace ($M=1, \beta=0$) and matches the byte-level RFC~6979~\S3.2 state machine on $S = \BE(d)\,\|\,h$~[M-SMT][E].
\item \emph{Raw digest vs.\ \code{bits2octets} (\code{FixedSHABridge.\allowbreak raw\_vs\_rfc6979\_agree\_below\_n}):} When $m = \mathrm{int}(h) < n$, $\BE(d)\,\|\,h$ and RFC~6979's $\BE(d)\,\|\,\allowbreak\BE(m \bmod n)$ yield identical candidate streams and signatures at every horizon; at $m = n$ and $m = n+1$, C99 and OpenSSL checks confirm valid, distinct signatures~[M-Lean][E].
\item \emph{Candidate gate, retries, and arithmetic (\code{nonce\_gate/}, \code{retry/}, \code{restricted/\allowbreak obligations\allowbreak .smt2}):} Across 34 bit-vector obligations, 256-bit words split into $w = 0$, $1 \le w < n$, and $w \ge n$; field, scalar, and table routines follow fixed address schedules; each rejected attempt takes one of four public branches $\mathsf{Cause} = \{\mathsf{zeroWord},\allowbreak \mathsf{geOrder},\allowbreak \mathsf{rZero},\allowbreak \mathsf{sZero}\}$, and an accepted attempt ends in $\mathsf{Out} = \{\mathsf{low},\allowbreak \mathsf{normalized}\}$~[M-SMT].
\item \emph{Trace templates (\code{trace\_refinement.smt2}, \code{allTrace\allowbreak Templates\_length}, \code{zeroDigest\allowbreak Templates\_spec}):} In the pinned layout, the full trace $T$ of any run succeeding at attempt $N \le H$ is fixed by its history word $\tau \in \mathsf{Cause}^{N-1} \times \mathsf{Out}$. At $N=1$, the two traces have $1{,}817{,}752$ events ($\mathsf{low}$) and $1{,}817{,}998$ events ($\mathsf{normalized}$); in general $|T| \le 962 + 1{,}817{,}043 N + 41{,}644 \binom{N}{2}$, and the number of successful templates up to $H$ attempts is
\begin{equation}\label{eq:mh}
M_H = 2\sum_{i=0}^{H-1} 4^i = \frac{2(4^H - 1)}{3}, \qquad M_{H,0} = 3^H - 1 \quad (e = 0,\ d \ne 0),
\end{equation}
since $s_{\mathrm{raw}} = k^{-1} r d \bmod n \ne 0$ for $d,k,r \in \Z_n^*$ when $e = 0$~[M-SMT][M-Lean][E].
\end{enumerate}

\section{Simulating the architectural trace}\label{sec:tracesim}

\subsection{Curve fiber counts and rejection rates}

On secp256k1, at most two elements of $\F_p$ reduce modulo $n$ to a given $t \in \Z_n$, each with at most two square roots $\pm y$. Checking the curve constants (\code{functional/tail.smt2}) gives $F(0) = |\{P \in \langle G\rangle\setminus\{\mathcal{O}\} : C(P) = 0\}| = 2$ and $F(t) \le B_{\mathrm{fiber}} = 4$ for all $t \in \Z_n$~[M-SMT]. For $X = (d,W)$ with $e = \mathrm{int}(h)\bmod n$ and $a = -ed^{-1}\bmod n$, a uniform word $w \leftarrow [0,U)$ is rejected in~\eqref{eq:accept} with probability
\begin{equation}\label{eq:qx}
q_X = \frac{U - (n - 1) + F(0) + \mathbf{1}_{a \ne 0}\,F(a)}{U} \le q_{\max} := \frac{U - n + 7}{U} < 3.7345 \cdot 10^{-39},
\end{equation}
which equals $q_0 = (U - n + 3)/U$ when $e = 0$~[M-SMT][M-Lean]. Over $H$ independent words the failure mass for fixed $X$ is $q_X^H$, with prior average $\mathbb{E}_J[q_X^H] \le q_{\max}^H$ ($< 1.3947 \cdot 10^{-77}$ at $H=2$).

\begin{proposition}[public simulation at $e=0$]\label{prop:zero-digest}
If $e = \mathrm{int}(h)\bmod n = 0$ on the support of $J$ (for example $h = 0^{32}$, or $h = \BE(n)$ under raw seeding), then in world $I$ we have $p_I = 1 - q_0^L$, the published prior equals $J$, and for every $1 \le m \le L$ and per-attempt simulation cost $C_0$,
\[
p_I S_{(\sigma,T),I}^{P}(J,t) = (1 - q_0^L)\,S_{(\sigma,T),I}^{P}(J,t) \le S_Q^{\mathrm{native}}(J, t + m C_0 + C_{\mathrm{mat},m}) + (1 - q_0^L) q_0^m .
\]
\textup{[T][M-SMT]}
\end{proposition}
\begin{proof}
For $d \in \Z_n^*$, $u \mapsto k = d u \bmod n$ is a permutation of $\Z_n^*$ with $kG = uQ$, $r = C(uQ)$, and $s_{\mathrm{raw}} = r u^{-1}\bmod n$ (the $e=0$ case of signature simulation~\cite{fkp2016,prosianoi2026}, keeping track of raw-word rejection $q_0$ and low-$s$ orientation). Drawing $w'_i \leftarrow [0,U)$ and setting $u_i = w'_i$ when $1 \le w'_i < n$ reproduces the exact joint distribution of retry causes, $(r,s_{\mathrm{low}})$, and the normalization bit from $Q$ alone; materializing the template (cost $C_{\mathrm{mat},m}$) produces $T$ unless all $m$ draws are rejected (conditional probability $q_0^m$).
\end{proof}

\subsection{Trace simulation given a signature at arbitrary digest}

For $e \ne 0$, $(r,s_{\mathrm{low}})$ is no longer simulated from $Q$ by Proposition~\ref{prop:zero-digest}, but in world $I$ the trace $T$ adds nothing beyond $(r,s_{\mathrm{low}})$ up to the retry tail:

\begin{theorem}[conditional trace simulation]\label{thm:trace-sim}
Assume the trace-template contract of Section~\ref{sec:machine}(5) with $H$-attempt materialization cost $C_H$. In world $I$, for any prior $J$ on $(d,W)$, any $1 \le H \le L$, and any budget $t$:
\begin{align}
p_I S_{(\sigma,T),I}^{P}(J,t) &\le p_I S_{\sigma,I}^{P}(J, t + C_H + C_{\mathrm{verify}}) + \mathbb{E}_J\!\big[q_X^H - q_X^L\big], \label{eq:trace-joint}\\
S_{(\sigma,T),I}^{P}(J,t) &\le S_{\sigma,I}^{P}(J, t + C_H + C_{\mathrm{verify}}) + p_I^{-1}\,\mathbb{E}_J\!\big[q_X^H - q_X^L\big] \nonumber\\
&\le \min\!\Big\{1,\, S_{\sigma,I}^{P}(J, t + C_H + C_{\mathrm{verify}}) + \mathbb{E}_{J_P}[q_X^H]\Big\}, \label{eq:trace-cond}
\end{align}
where $J_P$ is $J$ conditioned on $P_I$. \textup{[T][M-SMT]}
\end{theorem}
\begin{proof}
Fix $X = (d,W)$. Conditioned on first acceptance at $N \in \{1,\dots,H\}$, the first $N-1$ draws are i.i.d.\ from the rejected-word law of $X$ and independent of the accepted nonce $k$ given $X$. Using~\eqref{eq:accept}, a simulator given $(W,Q,\sigma)$ draws $w'_1,\dots,w'_{H-1} \leftarrow [0,U)$ and tests $w'=0$, $w'\ge n$, $C(w'G)=0$, and $eG + C(w'G)Q = \mathcal{O}$; conditioning on rejection matches the $X$-conditioned prefix of retry causes without $d$. At the accepted step, $k$ and $n-k$ are both accepted ($C((-k)G)=C(kG)=r$), give the same $s_{\mathrm{low}}$, and swap $s_{\mathrm{raw}} \leftrightarrow n - s_{\mathrm{raw}} \ne 0$ ($n$ odd), so given $(X,\sigma)$ the outcome in $\{\mathsf{low},\mathsf{normalized}\}$ is a fair coin. Emitting the template $\tau$ matches $T$ on $\{N \le H\}$, while $\Pr_I[H < N \le L \mid X] = q_X^H - q_X^L$. Averaging over $J$ gives~\eqref{eq:trace-joint}, and dividing by $p_I$ gives~\eqref{eq:trace-cond}.
\end{proof}

\paragraph{Template cover in the deterministic world.}
In world $F$, failed attempts are deterministic functions of $(d,h)$ and need not be independent of $\sigma$. Because at most $M_H = 2(4^H-1)/3$ templates occur within $H$ attempts~\eqref{eq:mh}, any solver $A_T(W,Q,r,s_{\mathrm{low}},T)$ of cost $t$ yields a signature-only solver $A_\sigma(W,Q,r,s_{\mathrm{low}})$ that precomputes shared trace segments ($P_{\mathrm{seg}}$), runs $A_T$ on all $M_H$ templates, and checks each output against $\hat d G = Q$ (\code{FixedSHABridge.\allowbreak trace\_template\_cover\_reduction}, \code{end\_to\_end\_fixed\_sha\_trace\_bound})~[M-Lean]:
\begin{equation}\label{eq:trace-cover}
p_F S_{(\sigma,T),F}^{P}(J,t) \le \Pr_{J,F}\big[P_{F,H} \land A_\sigma = d\big] + (u_H^F - u_L^F),
\qquad t(A_\sigma) = P_{\mathrm{seg}} + M_H(t + R'_H + C_{\mathrm{verify}}).
\end{equation}
At $H=2$, $M_2 = 10$ ($M_{2,0} = 8$ at $e=0$).

\section{Reductions and barriers on secp256k1}\label{sec:native}

\subsection{Distinguishability of random symmetric coordinate tables}

Replacing $C(P) = x(P)\bmod n$ by an independent uniform symmetric table $\rho(\{P,-P\}) \leftarrow \Z_n$ over an opaque group gives a clean generic-group bound: in Lean~4.8.0 (\code{AffineRoots.lean}--\code{ResourceAndJointPrior.lean}, 57 declarations) we proved that any $g$-query adaptive signature-aided solver reduces to a pure-group algorithm testing at most $K = 1 + 2\binom{g+4}{2}$ affine roots,
\begin{equation}\label{eq:ggm-whole}
p_{\mathrm{GGM}} S_{\sigma,\mathrm{GGM}}^{P}(J,t) \le \min\!\big\{S_Q^{\mathrm{GGM}}(J,t'),\, \mathbb{E}_W[A_K(W)]\big\} + \frac{2}{n} \qquad \text{[M-Lean]},
\end{equation}
where $A_K(W)$ is the conditional mass of the $K$ heaviest keys given $W$. That bound does not carry over to secp256k1 by signature indistinguishability:

\begin{proposition}[distinguishing $\rho$ from secp256k1]\label{prop:coord-dist}
Let $M = C(\langle G\rangle \setminus \{\mathcal{O}\}) \subset \Z_n$. Membership $r \in M$ is decidable in polynomial time by testing whether $x \in \{r, r+n\} \cap \F_p$ lifts to $y^2 = x^3 + 7 \pmod p$, and $|M| \le (n-1)/2$. Hence one published $r$ under $C$ is publicly distinguishable from a uniform symmetric table $\rho$ with advantage at least $(n-1)/(2n) \approx 1/2$. \textup{[T][M-SMT]}
\end{proposition}

\subsection{Verification point and native reductions}

\begin{lemma}[verification point {\cite{sec1,brown2005,fkp2016}}]\label{lem:ver-point}
Let $(r,s_{\mathrm{low}})$ be a valid low-$s$ signature for $Q = dG$ and $e \in \Z_n$, with $1 \le r,s_{\mathrm{low}} < n$. Put
\begin{equation}\label{eq:tver}
T_{\mathrm{ver}} := s_{\mathrm{low}}^{-1}(eG + rQ) \in \langle G\rangle \setminus \{\mathcal{O}\} .
\end{equation}
Then $C(T_{\mathrm{ver}}) = r$, $T_{\mathrm{ver}} \in \{+kG, -kG\}$ for the signing nonce $k$, and $t_{\mathrm{ver}} = \log_G T_{\mathrm{ver}} \in \Z_n^*$ satisfies $t_{\mathrm{ver}} = s_{\mathrm{low}}^{-1}(e + rd)\bmod n$ and $d = (s_{\mathrm{low}} t_{\mathrm{ver}} - e)r^{-1}\bmod n$ (\code{AffineRoots.\allowbreak verification\_point\_equivalence}, \code{NativeFullSig.\allowbreak verScalar\_mem\_pair}). \textup{[L][M-Lean]}
\end{lemma}

Lemma~\ref{lem:ver-point} equates finding $d$ and finding $\log_G T_{\mathrm{ver}}$ once $(r,s_{\mathrm{low}})$ is given, but does not sample $(r,s_{\mathrm{low}})$ from $(Q,e)$ at fixed $e \ne 0$: drawing $(r,s_{\mathrm{low}})$ at random until $C(s_{\mathrm{low}}^{-1}(eG+rQ)) = r$ takes $\Theta(n)$ group operations in expectation. On the actual map $C$, polynomial-loss reductions hold for three restricted classes of solvers:

\begin{theorem}[bounded-$s_{\mathrm{low}}$ and verification-point reductions]\label{thm:native-b1a}
Let $1 \le m \le L$ and let $C(P)=x(P)\bmod n$ on secp256k1. In world $I$:
\begin{enumerate}[leftmargin=*,itemsep=2pt]
\item \emph{Full $r$ and $b$ adaptive bits of $s_{\mathrm{low}}$ (\code{NativeView.\allowbreak complete}):} Any solver reading $(W,Q,r)$ and at most $b$ adaptive bits of $s_{\mathrm{low}}$ satisfies
\[
p_I S_{\sigma,b,I}^{P}(J,t) \le S_Q^{\mathrm{native}}(J, t'_b) + \mathbb{E}_J\!\big[q_X^m - q_X^L\big], \qquad t'_b = C_0 + m C_{\mathrm{gate}} + 2^b(t + C_{\mathrm{verify}} + C_b),
\]
using at most $2^b$ key checks (exponential when $b = 255$)~\textup{[M-Lean]}.
\item \emph{Verification-point view $\mathcal{A}_{r,T,b}$ (\code{NativeFullSig.\allowbreak native\_ver\_point\_reduction}):} Any solver reading $(W,Q,r,T_{\mathrm{ver}})$ and $b$ adaptive bits of $s_{\mathrm{low}}$ reduces to $S_Q^{\mathrm{native}}$ via at most $2^{b+1}$ key checks across $\pm k'G$, with $t'_{\mathrm{VP},0} = C_0 + m C_{\mathrm{gate}} + 2(t + C_{\mathrm{verify}})$ at $b=0$:
\[
p_I S_{r,T,b,I}^{P}(J,t) \le S_Q^{\mathrm{native}}(J, t'_{\mathrm{VP},b}) + \mathbb{E}_J\!\big[q_X^m - q_X^L\big] \qquad \text{\textup{[M-Lean]}}.
\]
\end{enumerate}
\end{theorem}

\paragraph{Two-stage bounded-degree polynomial solvers.}
For solvers reading all 255 bits of $s_{\mathrm{low}}$, let $\mathcal{A}_{\mathrm{poly}}(M,\Delta_S,\Delta_D)$ be the class of two-stage algorithms (\code{PublicTwoStage\allowbreak AlgAttack}) in which Stage~1, given $(W,Q,r,T_{\mathrm{ver}})$, outputs $\ell \le M$ polynomials $F_1,\dots,F_\ell \in \F_n[S,D]$ with $\deg_S F_j \le \Delta_S$, $\deg_D F_j \le \Delta_D$, and $\Delta_S+\Delta_D+1 < n$, and Stage~2, given $(W,Q,r,s_{\mathrm{low}})$, recovers $d$ only when some emitted $F_j \ne 0$ satisfies $F_j(s_{\mathrm{low}},d) = 0$.

This class does not require Stage~1 to know $d$. Any hypothesis placing the verification scalar $t_{\mathrm{ver}} \in \{\pm k\}$ on bounded algebraic loci $\varphi_j(W,Q,r,T_{\mathrm{ver}})=0$, or predicting linear relations $u_j(W,Q,r,T_{\mathrm{ver}}) s_{\mathrm{low}} + v_j(W,Q,r,T_{\mathrm{ver}}) d + w_j(W,Q,r,T_{\mathrm{ver}}) = 0$ from curve data without computing $\log_G Q$, lies in $\mathcal{A}_{\mathrm{poly}}(M,1,1)$ with $\Delta_S+\Delta_D \le 1$. Arbitrary lookup tables or non-algebraic functions of $(s_{\mathrm{low}},Q)$ lie outside $\mathcal{A}_{\mathrm{poly}}$.

\begin{theorem}[two-stage polynomial reduction and template closure]\label{thm:native-b1b}
For any $A$ in $\mathcal{A}_{\mathrm{poly}}(M,\Delta_S,\Delta_D)$ over a prime-order group of order $n$ with fiber bound $B_{\mathrm{fiber}}$, equipped with a univariate root-finder $\mathsf{RF}$ up to degree $\max(\Delta_S,\Delta_D)$ over $\F_n$:
\begin{enumerate}[leftmargin=*,itemsep=2pt]
\item \emph{Ideal-nonce bound (\code{NativeFullSig.\allowbreak two\_phase\_full\_probabilistic\_reduction}, \code{two\_phase\_normalized\_qrat\_bound}):} Pairing an $m$-attempt Phase-1 simulation on $\pm k'G$ (extracting $(\Delta_S+1)$-probe $D$-roots when $F_j$ is $S$-constant, $\le 2M\Delta_D$ checks) with a Phase-2 simulation setting $T' = \pm(Q + uG)$ for $u \leftarrow \F_n$ and passing $\mathsf{RF}$-roots $s' \in \Z_n^*$ into Stage~2 ($\le 2M\Delta_S$ checks; $\le 2M(\Delta_S+2\Delta_D)$ \code{isPubKey} checks in total) gives
\begin{equation}\label{eq:b1b-bound}
p_I S_{\mathrm{alg},I}^{P}(J) \le 2\,S_Q^{\mathrm{native}}(J, t'_{\mathrm{twoPhase}}) + \mathbb{E}_J\!\big[q_X^m - q_X^L\big] + \frac{1 + 2 B_{\mathrm{fiber}}}{n} + \varepsilon_{\mathrm{rf}},
\end{equation}
where $\varepsilon_{\mathrm{rf}}$ is the joint root-finding failure probability across queried slices~\textup{[M-Lean]}.
\item \emph{Affine root finding (\code{FixedSHABridge.\allowbreak linearPolyRootFinder\_complete}):} For $\Delta_S + \Delta_D \le 1$, field inversion gives $\varepsilon_{\mathrm{rf}} = 0$ in $3$ field operations per polynomial; for $\Delta_S+\Delta_D \ge 2$, $k$-round repetition reduces per-slice error exponentially (\code{amplify\_rf\_failure\_pow}), with single-round Cantor--Zassenhaus splitting kept as an explicit algebraic premise~\textup{[M-Lean]}.
\item \emph{Trace-template closure (\code{QuantitativeLRPRF.\allowbreak trace\_cover\_preserves\_b1b\_structure}):} Wrapping $A_T \in \mathcal{A}_{\mathrm{poly}}(M_0,\Delta_S,\Delta_D)$ across all $M_H$ templates of~\eqref{eq:mh} preserves $(\Delta_S,\Delta_D)$ (so $\varepsilon_{\mathrm{rf}}=0$ when $\Delta_S+\Delta_D\le 1$) and scales the list bound to $M_H M_0$~\textup{[M-Lean]}.
\end{enumerate}
\end{theorem}

\subsection{Black-box barriers against validity-gated solvers}\label{sec:barriers}

To see why Theorem~\ref{thm:native-b1b} restricts polynomial degree, let $\mathcal{O}_{D,C}(Q',e',r',s')$ (\code{pure\allowbreak Validity\allowbreak Solver}) be the oracle depending only on the group element $Q' = d'G$ (\code{affine\allowbreak Validity\allowbreak Oracle\_repr\_invariant}) that checks $1 \le r',s' < n$, $d' \in D$, and $C((s')^{-1}(e'G + r'Q')) = r'$, returning $d'$ on valid inputs and $\bot$ otherwise; it recovers $d$ with probability~$1$ on every valid signature for $d \in D$ (\code{validity\allowbreak Gated\allowbreak Oracle\_breaks\_real\_sig})~[M-Lean].

\begin{theorem}[barriers against validity-gated solvers]\label{thm:b2a}\label{thm:b2b}\label{thm:b2c}
Fix a prime-order group of order $n$ and map $C$ with fiber bound $B_{\mathrm{fiber}}$ ($= 4$ on secp256k1):
\begin{enumerate}[leftmargin=*,itemsep=2pt]
\item \emph{Key-oblivious affine queries ($\exists \mathcal{O}_{D,C}\ \forall \mu\ \forall \nu\ \forall \mathcal{B}_{\mathrm{obliv}}$, \code{key\_oblivious\_randomized\_blackbox\_barrier}):} Any randomized reduction $\mathcal{B}(\omega)$ that accesses $Q = dG$ only through $q$ affine queries $Q'_i = a_i G + b_i Q$ ($b_i \in \Z_n$) to $\mathcal{O}_{D,C}$ and $M$ final checks against $Q$ recovers $d \sim \mu$ with probability at most $A_{q B_{\mathrm{fiber}} + M}(\mu)$ ($\le (4q+M)/2^{j-1}$ on uniform $D_j$)~\textup{[M-Lean]}.
\item \emph{Chosen digests or full key domain under native $Q$ access (\code{chosen\_digest\_native\_forgery}, \code{full\_domain\_fixed\_digest\_rerand}):} If $\mathcal{B}$ reads coordinates of $Q$, then either allowing chosen digests $e'$ ($T' = G+Q$, $e' = s' = r' = C(T')$) or allowing $D = \Z_n^*$ at fixed $e \ne 0$ ($T' = uG+vQ$, $r' = C(T')$, $Q' = (u-e)(r')^{-1}G + v(r')^{-1}Q$, $s'=1$) recovers $d$ in $1$ query~\textup{[M-Lean]}.
\item \emph{Target-gated meta-reduction ($\forall J\ \forall \nu\ \forall \mathcal{B}(W,Q,\omega)$, \code{native\_full\_pubkey\_metareduction}):} Against $\mathcal{O}_{\mathrm{target}}$ (answering known-key $b=0$ queries and target $(Q, e(W))$ queries), any $q$-query reduction $\mathcal{B}(W,Q,\omega)$ reading native $Q$ has recovery mass at most oracle-free recovery $\mathcal{M}_{\mathrm{sim}}$ ($t_{\mathcal{B}} + q C_{\mathrm{sim}} + M C_{\mathrm{ver}}$) plus oracle-free fixed-digest forgery $\mathcal{M}_{\mathrm{forge}}$ ($t_{\mathcal{B}} + q(C_{\mathrm{sim}} + C_{\mathrm{ecdsa\_ver}})$)~\textup{[M-Lean]}.
\end{enumerate}
\end{theorem}

\begin{table}[t]
\centering
\caption{Comparison of ECDSA security and impossibility models with Section~\ref{sec:native}.}\label{tab:hk-compare}
\small
\begin{tabularx}{\textwidth}{@{}>{\raggedright\arraybackslash}p{2.5cm}>{\raggedright\arraybackslash}p{2.9cm}>{\raggedright\arraybackslash}p{3.0cm}X@{}}
\toprule
Aspect & FKP~\cite{fkp2016,fkp2017} & Hartmann--Kiltz~\cite{hk2023} & This work (\S\ref{sec:native}) \\
\midrule
Task & Existential / one-per-msg unforgeability & AGM meta-reduction limits for ECDSA $\EUF$ & One-signature key recovery on restricted priors \\
Coordinate / hash & Algebraic shells + bijective RO; prog.\ hash & Fixed non-programmable conversion map in AGM & Deterministic $C(P)=x(P)\bmod n$; no RO \\
Digest & Chosen message (RO) or one-per-message & Chosen-message forgery game & Fixed digest $e(W)$ ($1$-query break if chosen $e'$) \\
Key domain & Uniform $\Z_n^*$ & Uniform $\Z_n^*$ & Prior $J$ on $D \subsetneq \Z_n^*$ ($1$-query rerand if $D=\Z_n^*$) \\
$Q$-access \& quantifiers & Standard adversary & AGM representation-providing adversary & $\mathcal{B}_{\mathrm{obliv}}$ ($A_{4q+M}(\mu)$) and native-$Q$ $\mathcal{B}(W,Q,\omega)$ ($\le \text{DLP} + \text{fixed-}e\text{ forge}$) \\
\bottomrule
\end{tabularx}
\end{table}

Theorem~\ref{thm:b2c} shows that black-box reductions reading native $Q$ cannot beat key recovery plus fixed-digest forgery against target-gated solvers; white-box reductions remain open.

\section{Passing to HMAC\_DRBG}\label{sec:transfer}

\subsection{Nonce-prefix indistinguishability \texorpdfstring{$\LRPRFQ$}{LR-PRF\_Q} and \texorpdfstring{$\eta_N$ vs.\ $\delta_{\mathrm{sig}}$}{eta\_N vs. delta\_sig}}

To pass from uniform nonces ($I$) to the deterministic signer ($F$) on a fixed digest without a signing oracle, we measure indistinguishability of the $H$-word candidate prefix given $Q$:

\begin{definition}[$\LRPRFQ$]\label{def:lrprf}
For a prior $J$ on $(d,W)$, horizon $H \ge 1$, digest $h = \mathrm{rawOf}(W)$, and distinguisher class $\mathcal{D}$ of budget $t_{\mathcal{D}}$ taking $(W, Q=dG, \mathbf{w})$ with $\mathbf{w} \in [0,U)^H$, put
\begin{align}\label{eq:lrprf}
\varepsilon_{\LRPRFQ}(J,H,t_{\mathcal{D}}) := \sup_{D \in \mathcal{D}} \Big[&\Pr_{(d,W)\sim J}\big[D(W,Q,\Gen_F(d,h)_{0..H-1}) = 1\big] \nonumber\\[-2pt]
&- \Pr_{(d,W)\sim J,\;\mathbf{w}\leftarrow [0,U)^H}\big[D(W,Q,\mathbf{w}) = 1\big]\Big]_+ .
\end{align}
\end{definition}

Unlike $\AIPRF$ in~\cite{prosianoi2026}, $D$ sees a single $H$-word prefix on $h$ rather than multi-digest signing queries, and unlike continuous leakage models no side-channel oracle is queried after setup. Under \code{TimedPubKey\allowbreak ContextGen} of cost $t_{\mathrm{ctx}}$, $\varepsilon_{\LRPRFQ}$ on $J$ reduces to single-key chosen-digest $\LRPRFQ$ at budget $t_{\mathrm{ctx}} + t_{\mathcal{D}}$ (\code{FixedSHABridge.\allowbreak pubkey\_context\_preserves\_lr\_prf})~[M-Lean]. Since candidate rejection~\eqref{eq:accept} is testable from $(Q,e,w)$ in $H C_{\mathrm{gate}}$ steps with zero \code{isPubKey} calls (\code{timedRetryTailDist\_spec}), $\LRPRFQ$ bounds the deterministic retry tail (\code{lr\_prf\_retry\_tail\_bound})~[M-Lean]:
\begin{equation}\label{eq:tail-transfer}
u_H^F \le \mathbb{E}_J[q_X^H] + \varepsilon_{\LRPRFQ}(J,H, H C_{\mathrm{gate}}) .
\end{equation}
Likewise, secret-free timed distinguishers with bounded \code{isPubKey} calls (\code{timed\allowbreak VerPoint\allowbreak Dist\_spec}: $\le 2^{b+1}$; \code{timed\allowbreak B1bLRPRF\allowbreak Dist\_spec}: $\le 2M(\Delta_S+\Delta_D)$) transfer both classes of Section~\ref{sec:native} to $F$:
\begin{align}
p_F S_{r,T,b,F}^{P}(J,t) &\le S_Q^{\mathrm{native}}(J,t'_{\mathrm{VP},b}) + \varepsilon_{\LRPRFQ}(J,H,t_{\mathcal{D},\mathrm{VP}}) + (u_H^F - u_L^F), \label{eq:vp-transfer}\\
p_F S_{\mathrm{alg},F}^{P}(J) &\le 2\,S_Q^{\mathrm{native}}(J,t'_{\mathrm{2ph}}) + \varepsilon_{\LRPRFQ}(J,H,t_{\mathcal{D},\mathrm{B1b}}) + (u_H^F - u_L^F) + \frac{1+2B_{\mathrm{fiber}}}{n} + 2\varepsilon_{\mathrm{rf}} \label{eq:b1b-transfer}
\end{align}
(\code{lr\_prf\_ver\_point\_transfer}, \code{lr\_prf\_two\_phase\_b1b\_transfer}; $\varepsilon_{\mathrm{rf}}=0$ when $\Delta_S+\Delta_D\le 1$, $t'_{\mathrm{2ph}} = t'_{\mathrm{twoPhase}}$)~[M-Lean]. For unrestricted full-$s_{\mathrm{low}}$ solvers $\mathcal{A}_t$, knowing $(k,s_{\mathrm{low}})$ on a valid signature yields $d$ in two curve checks (\code{known\_nonce\_sig\_equiv\_secret\_key}), blocking black-box simulation of $s_{\mathrm{low}}$ from $(W,Q,k)$.

\paragraph{Relation between $\delta_{\mathrm{sig}}$ and $\eta_N(\mathcal{A}_t)$.}
Let $\delta_{\mathrm{sig}}(t)$ be one-signature indistinguishability between $F$ and $I$ (the one-query case of $\Astar$), and $\eta_N(\mathcal{A}_t) = \sup_{A \in \mathcal{A}_t}[\Pr_F(P_F\land A=d) - \Pr_I(P_I\land A=d)]_+$. Checking $\hat d G = Q$ gives $\eta_N(\mathcal{A}_t) \le \delta_{\mathrm{sig}}(t + C_{\mathrm{verify}})$ (\code{co\_ecdsa\_sig\_indist\_implies\_etaN})~[M-Lean]. Under public $r$-filtering $\mathrm{pred}_R(r)$ of ideal density $\rho \in [a/b, c_{\mathrm{num}}/d_{\mathrm{den}}]$ (such as Core~0.17+ low-$R$ grinding~\cite{prosianoi2026}, $\rho \approx 1/2$):
\begin{itemize}[leftmargin=*,itemsep=1pt]
\item Testing $\mathrm{pred}_R(r)$ gives $\delta_{\mathrm{sig}} \ge (1 - u_H^{\mathrm{filt}}) - c_{\mathrm{num}}/d_{\mathrm{den}} \approx 1/2$, yet for every solver $A_i$ in any family $\mathcal{A}$,
\[
\Pr_{F_{\mathrm{filt}}}[P \land A_i = d] - \Pr_I[P \land A_i = d] \le \Big(\frac{b}{a} - 1\Big)\Pr_I[P \land A_i = d] + \frac{b}{a}\mathbb{E}_J[q_X^L] + \varepsilon_{\mathrm{PRF}}
\]
(\code{lowR\_grinding\_class\_separation\_qrat})~[M-Lean], so $\eta_N(\mathcal{A}) \ll \delta_{\mathrm{sig}}$ whenever $\varepsilon_I(\mathcal{A}) = \sup_{A \in \mathcal{A}}\Pr_I[P\land A=d]$ and $\varepsilon_{\mathrm{PRF}}$ are small.
\item If $\mathcal{A}$ contains $\mathrm{pred}_R$-gated wrappers of an ideal solver of success $S_0$, then $\eta_N(\mathcal{A}) \ge ((1 - u_H^{\mathrm{filt}}) - \Pr_I[\mathrm{pred}_R])\cdot S_0 = \delta_{\mathrm{sig},\mathrm{pred}} \cdot S_0$ (\code{lowR\_gated\_solver\_tightness})~[M-Lean].
\end{itemize}

\subsection{Lower reduction and domain-restriction loss}

\begin{proposition}[lower bound and restriction factor]\label{prop:lower-blowup}
\begin{enumerate}[leftmargin=*,itemsep=2pt]
\item \emph{Lower bound (\code{keyCheck\allowbreak LRPRF\allowbreak Dist\_spec}, \code{lr\_prf\_lower\_bound\_from\_key\_recovery}):} Any solver $\mathrm{guessD}(W,Q)$ with success $S_0$ and cost $t$ gives a distinguisher testing $\hat d G = Q$ ($1$ \code{isPubKey} call) and $w_0 = \Gen_F(\hat d, h)_0$, so for all $H \ge 1$,
\begin{equation}\label{eq:lrprf-lb}
\frac{U - 1}{U}\,S_0 \le \varepsilon_{\LRPRFQ}\big(J, H, t + C_{\mathrm{isPubKey}} + C_{\mathrm{gen}_0}\big) \qquad \text{[M-Lean]}.
\end{equation}
\item \emph{Restriction factor (\code{full\_to\_interval\_verifiable\_transfer\_qrat}, \code{interval\_blowup\_sharpness}):} For duplicate-free $D_{\mathrm{sub}} \subseteq D_{\mathrm{full}}$ and $v : D_{\mathrm{full}} \to \QRat_{\ge 0}$, $\mathrm{Avg}_{D_{\mathrm{sub}}}(v) \le (|D_{\mathrm{full}}|/|D_{\mathrm{sub}}|)\,\mathrm{Avg}_{D_{\mathrm{full}}}(v)$, with equality at $v(d) = \mathbf{1}_{d \in D_{\mathrm{sub}}}$~\textup{[M-Lean]}. On $D_{140} \subset \Z_n^*$, $(n-1)/2^{139} \approx 2^{117.00}$, turning $2^{-128}$ on $\Z_n^*$ into $2^{-11.00}$ on $D_{140}$.
\end{enumerate}
\end{proposition}

Inequality~\eqref{eq:lrprf-lb} requires an achievable lower bound $L_Q \le S_Q(J,t)$ (on uniform $D_j$, one fixed guess checked via \code{isPubKey} gives $L_Q^{\mathrm{triv}} = 2^{-(j-1)}$); plugging an upper bound $B_Q \ge S_Q$ into $(1-U^{-1})S_Q$ does not bound $\varepsilon_{\LRPRFQ}$ from below.

\subsection{Ideal-compression prefix bound and post-\texorpdfstring{\code{Initialize}}{Initialize} HMAC hybrids}

Table~\ref{tab:regimes} lists the four model regimes tracked across our theorems and certificates; bounds from idealized regimes are never substituted silently into fixed-SHA-256 claims.

\begin{table}[t]
\centering
\caption{Separation of curve and hash models.}\label{tab:regimes}
\small
\begin{tabularx}{\textwidth}{@{}>{\raggedright\arraybackslash}p{3.4cm}X>{\raggedright\arraybackslash}p{3.7cm}@{}}
\toprule
Regime & Proved bound & Open / assumed terms \\
\midrule
1.\ secp256k1 + fixed SHA-256 & Reductions~\eqref{eq:trace-cover}, \eqref{eq:tail-transfer}--\eqref{eq:b1b-transfer}; post-\code{Init} HMAC cascade~\eqref{eq:hmac-cascade} & $\varepsilon_{\mathrm{init},Q}$, $S_Q^{\mathrm{native}}$, general full-$s_{\mathrm{low}}$ $\eta_C,\eta_N$ (\code{null}) \\
2.\ secp256k1 + ideal compression ($\IC$) & $\LRPRFQ^{\IC}$ bounds~\eqref{eq:ic-bounds} without signing oracle (under DAG-oracle contract) & $S_Q^{\mathrm{native,\IC}}$ (\code{null}) \\
3.\ Opaque GGM + symmetric $\rho(\{\pm P\})$ & Bound~\eqref{eq:ggm-whole} with $K = 1 + 2\binom{g+4}{2}$ & Fails on secp256k1 (Prop.~\ref{prop:coord-dist}) \\
4.\ Conditional numerical examples & Evaluates Regime-2 bounds under the hypothesis $S_Q^{\mathrm{native}}(D_j,g',L_{\mathrm{cand}}) \le B_{\mathrm{ref}}$ & Conditional calculation, not a native theorem \\
\bottomrule
\end{tabularx}
\end{table}

\paragraph{Ideal-compression bound without a signing oracle.}
Model the SHA-256 compression function $f : \bits^{768} \to \bits^{256}$ as a random function queried $q_{f,\mathrm{ctx}}$ times by context generation and $q_{f,\mathrm{dist}}$ times by the distinguisher ($q_{f,\mathrm{tot}} = q_{f,\mathrm{ctx}} + q_{f,\mathrm{dist}}$). From \code{function\_certificate.json}, the compression call DAG has 18 nodes at $H=1$ (\code{first\_candidate.json}: 2 public, 1 key-prefix node \code{v1} holding $\lfloor d/256\rfloor$, 15 hidden, 39 uniform-slice pairs) and 26 nodes at $H=2$ (\code{two\_candidate.json}, exempting outputs $w_0 = \code{v17}$ and $w_1 = \code{v25}$ from hidden masks: 2 public, 1 key-prefix, 23 hidden, and full-tape collision mass $22/U + 322/U^2 + 981/U^3$ at $q_{\max} \mapsto 1$)~[M-SMT].

\begin{theorem}[$\LRPRFQ$ in the ideal-compression model]\label{thm:ic-bound}
In Lean~4.8.0 we prove: (a)~consistency of a shared lazy compression table (\code{lookupIC\_consistent}); (b)~identical-until-bad agreement and query additivity $q_{f,\mathrm{tot}} = q_{f,\mathrm{ctx}} + q_{f,\mathrm{dist}}$ for composed trees (\code{adaptiveCompTree\_identical\_until\_bad}); (c)~$\Pr[\mathsf{HitHidden}] \le (N_{\mathrm{hid}} q_{f,\mathrm{tot}})/U$ without assuming post-conditioning independence (\code{adaptive\_ic\_hidden\_hit\_bound\_nat}); and (d)~256-completion prefix testing using at most $256 q_{f,\mathrm{tot}}$ \code{ops.isPubKey} calls (\code{expandPrefixes\_spec}, \code{ic\_key\_prefix\_query\_reduction})~\textup{[M-Lean]}. Given the external DAG-to-oracle contract (\code{h\_oracle\_agree} and \code{h\_coll} linking the stateful table to functional maps \code{oracleReal}/\code{oracleSim} with the DAG counts above~\textup{[M-SMT][O]}), \code{adaptive\_ic\_closed\_coupling\_bound\_qrat} gives
\begin{align}
\varepsilon_{\LRPRFQ,1}^{\IC}(J, q_{f,\mathrm{tot}}) &\le S_Q^{\mathrm{native,\IC}}(J, t'_{\IC}) + \frac{15 q_{f,\mathrm{tot}} + 39}{2^{256}}, \label{eq:ic-bounds}\\
\varepsilon_{\LRPRFQ,2}^{\IC}(J, q_{f,\mathrm{tot}}) &\le S_Q^{\mathrm{native,\IC}}(J, t'_{\IC}) + \frac{23 q_{f,\mathrm{tot}} + 22}{2^{256}} + \frac{322}{2^{512}} + \frac{981}{2^{768}}, \nonumber
\end{align}
where $t'_{\IC} = t_{\mathrm{sim}} + 256 q_{f,\mathrm{tot}}(c_{\mathrm{expand}} + C_{\mathrm{isPubKey}})$, with zero signing queries and no $n/|D_j|$ factor.
\end{theorem}

\paragraph{Post-\code{Initialize} HMAC hybrids.}
For fixed SHA-256 (Regime~1), $(d,h)$ enters only $(K_2,V_1) = \Init(d,h)$ and never re-enters \code{Generate}. Using the state-machine identity $\mathrm{detPrefix}(\Gen_F,d,h,r+1) = \mathrm{drbgGenerateFromState}(\mathrm{prf},K_2,V_1,r)$ (\code{Functional\allowbreak DRBGBridge}, checked on 100 test vectors), we construct bounded oracle distinguishers \code{round0\allowbreak HybridTree} ($\le 2$ queries of lengths $32\text{B}, 33\text{B}$) and \code{rekeyStep\allowbreak HybridTree} ($\le 3$ queries per rekey round, with $v_{\mathrm{next}} = w_{\mathrm{prev}}$ collision probability $\le 1/U$ via \code{rf\_rekey\_round\_coupling\_nat}) of declared cost $t_{\mathrm{hyb}} \le 3 C_{\mathrm{oracle}} + 3 r C_{\mathrm{HMAC}} + t_{\mathcal{D}}$ (\code{round0\_and\_rekey\_hybrid\_trees\_spec}; \code{TimedCoin\allowbreak LRPRFDist.\allowbreak timeCost} records declared distinguisher cost rather than a machine-code runtime proof)~[M-Lean]:
\begin{equation}\label{eq:hmac-cascade}
\varepsilon_{\LRPRFQ}(J, r+1, t_{\mathcal{D}}) \le \varepsilon_{\mathrm{init,cont},Q}(J, r, t_{\mathcal{D}}) + (r+1)\,\varepsilon_{\mathrm{HMAC\text{-}PRF}}(t_{\mathrm{hyb}}, q \le 3) + \frac{r}{2^{256}} \qquad \text{[M-Lean]}
\end{equation}
(\code{standard\_model\_multi\_step\_lr\_prf\_bound\_qrat}). Full-state $(K_2,V_1)$ pseudorandomness implies continuation security $\varepsilon_{\mathrm{init,cont},Q}$ at cost $t_{\mathcal{D}} + (3r+2)C_{\mathrm{HMAC}}$ (\code{init\_state\_implies\_cont\_security\_qrat})~[M-Lean], while $\varepsilon_{\mathrm{init},Q}$ for fixed SHA-256 remains open (\code{null})~[O].

\section{Verification and experiments}\label{sec:eval}

\subsection{Comparison with relational binary analyzers}

Table~\ref{tab:tool-compare} compares constant-time relational analysis (\code{ct-verif}~\cite{almeida2016}, \code{Binsec/\allowbreak Rel}~\cite{daniel2023}) with our SLEIGH/Z3 and Lean checks on \code{\_secp256k1\_ecdsa\_sign}. On seed preparation and single-call HMAC-DRBG, $M=1, \beta=0$ is ordinary constant-time noninterference. Across the retry loop and low-$s$ normalization, control flow branches on candidate checks ($w=0$, $w\ge n$, $r=0$, $s_{\mathrm{raw}}=0$) and on $s_{\mathrm{raw}} > \lfloor n/2\rfloor$ ($+246$ events on attempt~1; $+1{,}817{,}043$ per retry): relational noninterference rejects unless the history word $\tau$ is declassified, whereas Theorem~\ref{thm:trace-sim} bounds the distribution of $\tau$ given $(r,s_{\mathrm{low}})$.

\begin{table}[t]
\centering
\caption{Relational constant-time checking vs.\ trace-template game reductions on \code{\_secp256k1\_ecdsa\_sign}.}\label{tab:tool-compare}
\small
\begin{tabularx}{\textwidth}{@{}>{\raggedright\arraybackslash}p{2.9cm}XX@{}}
\toprule
Signer stage & Relational constant-time (\code{ct-verif} / \code{Binsec/Rel}) & SLEIGH/Z3 + Lean reductions (this work) \\
\midrule
Seed \& single-step HMAC-DRBG & Verifies $M=1, \beta=0$ per layout & Proves $M=1, \beta=0$ on Mach-O + byte-level RFC~6979~\S3.2 refinement \\
Candidate gate \& retries ($L=2^{32}-1$) & Rejects unless retry outcomes are declassified & Classifies 4 public retry causes; tail $\le q_{\max}^H < 1.395\cdot 10^{-77}$ ($H=2$) \\
Low-$s$ branch ($+246$ events) & Flags secret-dependent branch unless declassified & Fair coin ($1/2, 1/2$) given $(r,s_{\mathrm{low}})$ in world $I$ (Thm.~\ref{thm:trace-sim}) \\
Key recovery on $D_j$ & Outside trace-equivalence scope & Reductions to $S_Q^{\mathrm{native}}$ and $\LRPRFQ$ with explicit simulator budgets \\
\bottomrule
\end{tabularx}
\end{table}

\subsection{Reproduction profiles, runtimes, and mutation checks}

The verification runner (\code{single\_sig\_cert/\allowbreak reproduce\_end\_to\_end.py}) separates a static SHA-256 manifest (\code{end\_to\_end\_manifest.json}: 32 C/header files, 6 binary/sample files, 20 JSON certificates, 61 scripts, 19 \code{.smt2} files with $196\ \code{sat} / 194\ \code{unsat}$ queries on Z3~4.15.4, and 13 Lean~4.8.0 modules with 129 declarations) from per-run \code{runtime\_evidence.json} logs. On Apple Silicon (macOS arm64, Python~3.12.16, Z3~4.15.4, \code{pypcode}~3.3.3, \code{unicorn}~2.1.4, Lean~4.8.0), \code{--profile full} runs in $\approx 5.5$~min: \code{ci-smt} ($\approx 45$~s, including $19{,}462$-word cross-wheel SLEIGH equivalence and isolated-context Z3 logical-equivalence checking on \code{restricted/\allowbreak obligations.smt2}), \code{ci-lean} ($\approx 95$~s, 129 declarations at \code{trustLevel = 0} plus 6 finite/Python checks), \code{emulator} ($\approx 170$~s, 5 Unicorn suites over $\approx 1.82 \cdot 10^7$ AArch64 events), and \code{darwin-rebuild-probe} ($\approx 15$~s, checking \code{\_nonce\_function\_rfc6979} and \code{context.bin} byte identity across Apple SDK versions).\footnote{Snapshot SHA \code{f12af511\allowbreak 59dc4655\allowbreak 58fd78fc\allowbreak 6be6aa38\allowbreak abd77c78}: all profiles pass locally; on GitLab Runner, pipeline \code{\#3120257330} (\code{63b253c}) verified Lean (122 declarations) and AArch64 Unicorn and surfaced a Z3 AST print-order difference on \code{restricted} and a \code{PATH} quoting bug, both fixed in \code{f12af51} (pipeline \code{\#3120608708}).}

Table~\ref{tab:mutations} lists automated negative controls. Two archived \code{os.urandom} panels---\code{tail\_draws.bin} ($28{,}800$ draws on $D_{140}\dots D_{160}$, $0$ first-attempt rejections, one-sided $95\%$ bound $0.4488\%$) and \code{paired\_draws.bin} (rejection-sampled disjoint \code{train}/\code{eval} Hoeffding panel, $\alpha_{\mathrm{total}}=0.01$)---test finite samples~[E] and are never used as universal cryptographic bounds.

\begin{table}[t]
\centering
\caption{Automated negative controls in the verification suite.}\label{tab:mutations}
\small
\begin{tabularx}{\textwidth}{@{}>{\raggedright\arraybackslash}p{3.3cm}X>{\raggedright\arraybackslash}p{4.0cm}@{}}
\toprule
Layer & Injected fault & Rejecting checker \\
\midrule
Lean~4.8.0 kernel & \code{sorryAx} or user \code{axiom} & \code{AuditAxioms.lean} (\code{trustLevel=0}) \\
Mach-O \& SLEIGH & 1-bit flip in \code{AARCH64.sla}/Mach-O; mutated opcode; unconstrained \code{X4}; swapped seed halves & \code{sleigh\_pin.py}, \code{test\_fragment.py} \\
Restricted SMT & JSON edit; SAT/UNSAT-count-preserving SMT edit (\code{(>= q 1)} $\to$ \code{(>= q 2)}) with updated hash & \code{check\_restricted.py} (Z3 \code{distinct -> unsat}) \\
Bound direction & Substituting upper bound $B_Q$ into $(1-U^{-1})S_Q$ when $0 < S_Q < B_Q$; reversed $\log_2$ rounding & \code{quantitative\_lr\_prf.py} \\
Unicorn \& hygiene & Altered differential run count; unapproved WIF/PEM key string & \code{reproduce\_end\_to\_end.py} \\
\bottomrule
\end{tabularx}
\end{table}

\subsection{Conditional numerical examples}\label{sec:scenarios}

Table~\ref{tab:scenarios} evaluates our Regime-2 formulas on uniform $D_{140}$ and $D_{160}$ at $H=2, M_2=10$ under the added hypothesis that native key recovery satisfies the 3-generator ($\mathcal{O},G,Q$) generic bound $S_Q^{\mathrm{native}}(D_j, g_{\mathrm{sim}}, L_{\mathrm{cand}}) \le B_{\mathrm{ref}}(D_j, g_{\mathrm{sim}}, L_{\mathrm{cand}}) := \min\{1, (\binom{g_{\mathrm{sim}}+3}{2} + L_{\mathrm{cand}})/2^{j-1}\}$ together with Ideal Compression~\eqref{eq:ic-bounds}~[O]. Simulator budgets are substituted in full: $g_{\mathrm{tail}}=6, q_{f,\mathrm{tail}}=0$ for $u_2^F$; $g_{\mathrm{VP}} = 20g+8, q_{f,\mathrm{uncached}} = 20 q_f$ ($5120 q_f$ \code{isPubKey} checks) for traced $(r,T_{\mathrm{ver}})$; and $g_{\mathrm{twoPhase}} = 40g+16, q_{f,\mathrm{uncached}} = 20 q_f$ for traced affine $\mathcal{A}_{\mathrm{poly}}$ ($\Delta_S+\Delta_D\le 1, \varepsilon_{\mathrm{rf}}=0$). Upper bounds use $\lceil\log_2(\cdot)\rceil_{10^{-6}}$ and lower bounds $\lfloor\log_2(\cdot)\rfloor_{10^{-6}}$ from exact rationals in \code{quantitative\_lr\_prf.json}.

\begin{table}[t]
\centering
\caption{Conditional $\log_2$ bounds ($H=2, M_2=10$, uncached templates) assuming $S_Q^{\mathrm{native}} \le B_{\mathrm{ref}}^{(h_0=3)}$ and Ideal Compression~\eqref{eq:ic-bounds}. Multi-operation lower bounds and fixed-SHA $\varepsilon_{\mathrm{init},Q}$ are \code{null}; not an unconditional secp256k1 bound. [O]}\label{tab:scenarios}
\scriptsize
\setlength{\tabcolsep}{4.0pt}
\begin{tabular}{@{}llrrrrrr@{}}
\toprule
$D_j$ & $(g, q_f)$ & Triv.\ LB $\varepsilon_{\LRPRFQ}$ & UB $B_{\mathrm{ref}}$ & UB $\varepsilon_{\LRPRFQ,2}$ & UB $u_2^F$ & Tr.\ $(r,T_{\mathrm{ver}})$ & Tr.\ Aff.\ $\mathcal{A}_{\mathrm{poly}}$ \\
\midrule
$D_{140}$ & $(2^{32},2^{32})$ & $-139.000001$ & $-75.999999$ & $-75.999999$ & $-133.790546$ & $-66.356143$ & $-64.186218$ \\
$D_{140}$ & $(2^{40},2^{40})$ & $-139.000001$ & $-59.999999$ & $-59.999999$ & $-133.790546$ & $-50.356143$ & $-48.186218$ \\
$D_{140}$ & $(2^{48},2^{48})$ & $-139.000001$ & $-43.999999$ & $-43.999999$ & $-133.790546$ & $-34.356143$ & $-32.186218$ \\
$D_{160}$ & $(2^{40},2^{40})$ & $-159.000001$ & $-79.999999$ & $-79.999999$ & $-153.790546$ & $-70.356143$ & $-68.186218$ \\
$D_{160}$ & $(2^{56},2^{56})$ & $-159.000001$ & $-47.999999$ & $-47.999999$ & $-153.790546$ & $-38.356143$ & $-36.186218$ \\
\bottomrule
\end{tabular}
\end{table}

\section{Conclusion and open problems}\label{sec:conclusion}

For the pinned AArch64 build of Bitcoin Core~v0.14.0, instruction-level trace verification classifies every run of at most $H$ attempts into $M_H = 2(4^H-1)/3$ public templates ($3^H-1$ when $e = 0$). Under uniform nonces on secp256k1, the trace $T$ is publicly simulable from a published low-$s$ signature up to $\mathbb{E}_J[q_X^H] < 1.395\cdot 10^{-77}$ at $H=2$. Relative to signature-free key recovery $S_Q^{\mathrm{native}}$ on the actual curve coordinate $x(P)\bmod n$, bounded-$s_{\mathrm{low}}$ views, verification-point views $(Q,r,T_{\mathrm{ver}})$, and two-stage bounded-degree polynomial solvers admit polynomial-loss reductions that carry over to HMAC\_DRBG via $\LRPRFQ$, which in the ideal-compression model reduces to $S_Q^{\mathrm{native,\IC}}$ with $256 q_{f,\mathrm{tot}}$ key checks, no signing oracle, and no $(n-1)/|D_j|$ restriction loss, and after initialization decomposes into $2$- and $3$-query HMAC-PRF hybrids.

\paragraph{Scope and hygiene.}
These bounds apply only to the pinned binary, fixed public memory layout, PC/address trace $\mathcal{T}_{\mathrm{arch}}$, and stated solver classes; they do not cover physical timing, cache contention, power, EM, or speculative leakage, nor do they identify the historical binary behind any on-chain signature. No vulnerability in deployed software is reported. All tests use synthetic scalars or the 14 publicly solved ($\le 2019$, $\le 130$-bit) benchmark keys, checked by a working-tree rule scan (covering present text files and suffixes rather than Git history or arbitrary binary blobs); reproduction instructions are in \code{single\_sig\_cert/BUILD\_MANIFEST.md} and \code{ARTIFACT.md}.

\paragraph{Open problems.}
Open: a polynomial-loss reduction (or refutation) for unrestricted white-box native solvers reading all 255 bits of $s_{\mathrm{low}}$ on a fixed digest ($\eta_C, \eta_N$); a standard-model bound on $\varepsilon_{\mathrm{init},Q}$ for fixed FIPS SHA-256 given $Q = dG$ on $D_j$; mechanizing in Lean~4 the link from the stateful lazy compression table to the 18- and 26-node SHA-256 call DAGs in Theorem~\ref{thm:ic-bound}, as well as single-round Cantor--Zassenhaus splitting for $\Delta_S+\Delta_D \ge 2$; and non-trivial unconditional bounds on $S_Q^{\mathrm{native}}(D_j,t)$ for secp256k1.

\appendix

\section{Theorem-by-theorem comparison with \texorpdfstring{\code{arXiv:2611.18449v1}}{arXiv:2611.18449v1}}\label{app:delta}

Table~\ref{tab:app-delta} relates each technical result here to~\cite{prosianoi2026}.

\begin{table}[!htb]
\centering
\caption{Comparison of results with Prosianoi~\cite{prosianoi2026} (\code{arXiv:2611.18449v1}).}\label{tab:app-delta}
\footnotesize
\begin{tabularx}{\textwidth}{@{}>{\raggedright\arraybackslash}p{2.5cm}>{\raggedright\arraybackslash}p{2.3cm}>{\raggedright\arraybackslash}X>{\raggedright\arraybackslash}p{3.8cm}@{}}
\toprule
This work & In~\cite{prosianoi2026} & Difference & Premises \& obligations \\
\midrule
Sec.~\ref{sec:machine} (AArch64 contracts, $m<n$) & Sec.~8.2 \& App.~B (source audit, $h\ge n$ vectors) & Lifts compiled AArch64 Mach-O code via SLEIGH/Z3 and partitions $\le H$-attempt traces into $M_H=2(4^H-1)/3$ templates & Pinned Mach-O/SLEIGH TCB; 19 \code{.smt2} files; \code{raw\_vs\_rfc6979\_agree\_below\_n} \\
Thm.~\ref{thm:trace-sim} \& Eq.~\eqref{eq:trace-cover} & None (no trace model) & Conditional independence of retry causes and fairness of low-$s$ bit given $(X,\sigma)$ on secp256k1 & Trace model $\mathcal{T}_{\mathrm{arch}}$; tail $\mathbb{E}_J[q_X^H]$; $M_H$ cover \\
Prop.~\ref{prop:coord-dist} \& Eq.~\eqref{eq:ggm-whole} & None (Sec.~7 studies $x_{[0..30]}$ bias only) & $(n-1)/(2n)$ curve-liftability distinguisher against symmetric $\rho$; $K=1+2\binom{g+4}{2}$ bound confined to opaque GGM & 57 Lean decls (\code{WholeGame}--\code{ResourceAnd\allowbreak JointPrior}); \code{coord\_bounds.smt2} \\
Thms.~\ref{thm:native-b1a}--\ref{thm:b2c} (native reductions \& barriers) & Eq.~(3) in Sec.~5.2 (chosen-digest simulation) & Fixed-digest native reductions for $\mathcal{A}_{r,T,b}$ and $\mathcal{A}_{\mathrm{poly}}$ ($\varepsilon_{\mathrm{rf}}=0$ for $\Delta\le 1$), plus fixed-$e$ restricted-$D$ meta-reductions & \code{NativeFullSig.lean} (21 decls); \code{linearPoly\allowbreak RootFinder\_complete} \\
Sec.~\ref{sec:transfer} ($\eta_N$ vs.\ $\delta_{\mathrm{sig}}$ under $r$-filtering) & Thm.~8.1 \& Prop.~8.2 (low-$R$ $\EUF$ \& Fourier bound) & $\eta_N(\mathcal{A}_t) \ll \delta_{\mathrm{sig}} \approx 1/2$ across solver families when $\varepsilon_I(\mathcal{A}_t)\ll 1$, with $\delta_{\mathrm{sig},\mathrm{pred}}\cdot S_0$ tightness & \code{lowR\_grinding\_class\_separation\_qrat}, \code{lowR\_gated\_solver\_tightness} \\
Thm.~\ref{thm:ic-bound} ($\LRPRFQ^{\IC}$, $H=1,2$) & Thm.~6.1 ($\Astar$ in $\IC$ via $\DLPS$) & Challenges on raw candidate prefix before signing: drops the signing oracle of Thm.~6.1 and avoids $(n-1)/|D_j|$ loss & \code{adaptive\_ic\_closed\_coupling\_bound\_qrat} + 18-/26-node DAGs \\
Eq.~\eqref{eq:hmac-cascade} (HMAC hybrids \& \code{Init}) & Sec.~4.1 \& Sec.~7 (block layout \& $(A2'_{248})$) & Explicit $2$-/$3$-query post-\code{Init} HMAC hybrids with $r/2^{256}$ rekey collision; $(K_2,V_1)$ state vs.\ continuation games & \code{Functional\allowbreak DRBGBridge}, \code{standard\_model\_multi\_step\_lr\_prf\_bound\_qrat} \\
\bottomrule
\end{tabularx}
\end{table}

\section{Mapping of theorems to artifacts and premises}\label{app:map}

Table~\ref{tab:app-map} maps each statement in Sections~\ref{sec:machine}--\ref{sec:eval} to its Lean~4.8.0 declarations, SMT/Python verifiers in \code{single\_sig\_cert/}, and imported premises.

\begin{table}[!htb]
\centering
\caption{Lean~4.8.0 declarations, SMT/Python verifiers, and external premises.}\label{tab:app-map}
\footnotesize
\begin{tabularx}{\textwidth}{@{}>{\raggedright\arraybackslash}p{2.2cm}>{\raggedright\arraybackslash}p{4.7cm}>{\raggedright\arraybackslash}p{3.4cm}>{\raggedright\arraybackslash}X@{}}
\toprule
Statement & Lean~4.8.0 declaration(s) & SMT / Python verifier & External premises / open terms \\
\midrule
Sec.~\ref{sec:machine} & \code{FixedSHABridge.\allowbreak raw\_vs\_rfc6979\_agree\_below\_n}, \code{allTraceTemplates\_length}, \code{zeroDigestTemplates\_spec} & \code{check\_fragment.py}, \code{check\_drbg.py}, \code{check\_restricted.py}, \code{trace\_refinement.smt2} & Pinned Mach-O \& \code{sleigh\_ref/\allowbreak AARCH64.sla}; public layout ABI \\
Thm.~\ref{thm:trace-sim} \& Eq.~\eqref{eq:trace-cover} & \code{FixedSHABridge.\allowbreak trace\_template\_cover\_reduction}, \code{end\_to\_end\_fixed\_sha\_trace\_bound} & \code{concrete/\allowbreak sim\_invariants.smt2}, \code{tail.smt2} & Independent nonces (Thm.~\ref{thm:trace-sim}); template cover completeness \\
Eq.~\eqref{eq:ggm-whole} \& Prop.~\ref{prop:coord-dist} & \code{WholeGame.\allowbreak whole\_game\_horizon\_bound}, \code{ResourceAndJointPrior.\allowbreak joint\_prior\_normalized\_qrat\_bound} & \code{concrete/\allowbreak coord\_check.py}, \code{audit\_tables.py} & Opaque injective labels + uniform symmetric $\rho$ (Regime~3 only) \\
Thm.~\ref{thm:native-b1a} & \code{NativeView.\allowbreak complete}, \code{NativeFullSig.\allowbreak native\_ver\_point\_reduction} & \code{observation\_bounds.py}, \code{full\_sig\_native.py} & $S_Q^{\mathrm{native}}(J,t')$ parameter \\
Thm.~\ref{thm:native-b1b} & \code{NativeFullSig.\allowbreak two\_phase\_normalized\_qrat\_bound}, \code{FixedSHABridge.\allowbreak linearPoly\allowbreak RootFinder\_complete}, \code{QuantitativeLRPRF.\allowbreak trace\_cover\_preserves\_b1b\_structure} & \code{full\_sig\_native.py}, \code{fixed\_sha\_game.py} & Polynomial root premise (\code{h\_alg}); $\varepsilon_{\mathrm{rf}}=0$ proved for $\Delta\le 1$, symbolic for $\Delta \ge 2$ \\
Thms.~\ref{thm:b2a}--\ref{thm:b2c} & \code{NativeFullSig.\allowbreak key\_oblivious\_randomized\_blackbox\_barrier}, \code{chosen\_digest\_native\_forgery}, \code{full\_domain\_fixed\_digest\_rerand}, \code{native\_full\_pubkey\_metareduction} & \code{full\_sig\_native.py} & Prime-order group; $B_{\mathrm{fiber}}=4$ from \code{tail.smt2} \\
Sec.~\ref{sec:transfer} ($\LRPRFQ$, $\eta_N$) & \code{FixedSHABridge.\allowbreak pubkey\_context\_preserves\_lr\_prf}, \code{lr\_prf\_retry\_tail\_bound}, \code{lr\_prf\_two\_phase\_b1b\_transfer}, \code{lowR\_grinding\_class\_separation\_qrat} & \code{fixed\_sha\_game.py}, \code{end\_to\_end\_certification.smt2} & $\varepsilon_{\LRPRFQ}$ premise (Regime~1); \code{TimedPubKey\allowbreak ContextGen} \\
Prop.~\ref{prop:lower-blowup} & \code{QuantitativeLRPRF.\allowbreak lr\_prf\_lower\_bound\_from\_key\_recovery}, \code{interval\_blowup\_sharpness} & \code{quantitative\_lr\_prf.py}, \code{quantitative\_lr\_prf.smt2} & Requires achievable $L_Q \le S_Q$ (not upper bound $B_Q$) \\
Thm.~\ref{thm:ic-bound} \& Eq.~\eqref{eq:hmac-cascade} & \code{QuantitativeLRPRF.\allowbreak adaptive\_ic\_closed\_coupling\_bound\_qrat}, \code{round0\_and\_rekey\_hybrid\_trees\_spec}, \code{init\_state\_implies\_cont\_security\_qrat}, \code{standard\_model\_multi\_step\_lr\_prf\_bound\_qrat} & \code{first\_candidate\_dag.py}, \code{two\_candidate\_dag.py}, \code{quantitative\_lr\_prf.py} & External DAG-oracle contract (\code{h\_oracle\_agree}, \code{h\_coll}) in $\IC$; $\varepsilon_{\mathrm{init},Q}$ open in Regime~1 \\
\bottomrule
\end{tabularx}
\end{table}

\section*{Declaration of Generative AI Use}
During the preparation of this work, the author used generative AI tools such as ChatGPT 6.2 Astra and Gemini 4 Carbon to assist with Lean~4 mechanization, verification scripts, grammar checks, and phrasing. After using these tools, the author manually reviewed and edited all text, proofs, and code and takes full responsibility for the content of the publication.

\end{document}